\documentclass[journal]{IEEEtran}
\usepackage{upgreek}
\usepackage{graphicx}
\usepackage{subfigure}
\usepackage{amsmath,bm}
\usepackage{amssymb}

\usepackage{pifont}
\usepackage{enumerate}

\usepackage{algorithm}
\usepackage{algpseudocode}

\usepackage{url}

\usepackage[square, comma, sort&compress, numbers]{natbib}

\usepackage{color}
\newcommand{\mll}{\color{black}}
\newcommand{\mrr}{\color{black}}

\begin{document}

\title{On the Stability of Spatially Distributed Cavity Laser and Boundary of Resonant Beam SLIPT}
\author{
	Mingliang Xiong,~\IEEEmembership{Member,~IEEE},  
    Zeqian Guo,
    Qingqing Zhang,~\IEEEmembership{Member,~IEEE}, Qingwen Liu,~\IEEEmembership{Senior Member,~IEEE},  Gang Wang,~\IEEEmembership{Senior Member,~IEEE}, Gang Li,~\IEEEmembership{Member,~IEEE}, and Bin He,~\IEEEmembership{Senior Member,~IEEE}
\thanks{

}
\thanks{This work was supported in part by the National Natural Science Foundation of China under Grant 62305019 and U23B2059. } 
\thanks{
	M. Xiong, Z. Guo and Q. Liu
	are with the School of Computer Science and Technology, Tongji University, Shanghai 201804, China (e-mail: mlx@tongji.edu.cn; guozeqian@tongji.edu.cn; qliu@tongji.edu.cn)
	
	Q. Zhang is with the College of Information Engineering, Zhejiang University of Technology, Hangzhou 310014, China (e-mail: qingqingzhang@zjut.edu.cn)

    G. Wang is with the State Key Lab of Intelligent Autonomous Systems, School of Automation, Beijing Institute of Technology, Beijing 100081, China (e-mail: gangwang@bit.edu.cn)
    
    G. Li and B. He are with  Shanghai Research Institute for Autonomous Intelligent Systems, Tongji University, Shanghai 201804, China. (e-mail: lig@tongji.edu.cn; hebin@tongji.edu.cn)}
}

\maketitle

\begin{abstract}
Spatially distributed cavity (SDC) lasers are a promising technology for simultaneous light information and power transfer (SLIPT), offering benefits such as increased mobility and intrinsic safety, which are advantageous for various Internet of Things (IoT) devices. \mll However, achieving beam transmission over meter-level long working distances presents significant challenges from cavity stability constraints, manufacturing/assembly tolerances, and diffraction losses\mrr. This paper conducts a theoretical investigation of the fundamental restrictions limiting long-range resonant beam generation. We investigate cavity stability and beam characteristics, and propose a binary-search-based Monte Carlo simulation algorithm as well as a linear approximation algorithm to quantify the maximum acceptable tolerances for stable operation. \mll Numerical results indicate that the stable region contracts sharply as distance increases. For fixed-component systems, an acceptable tolerance of $0.01$ mm restricts the achievable transmission distance to less than 2 m. \mrr To address this limitation, we also prove the feasibility of long-range beam formation using precision adjustable elements, paving the way for advanced engineering applications. \mll Experimental results verified this assumption, demonstrating that by tuning the stable region during assembly, the transmission distance could be extended to $2.8$ m. \mrr This work provides essential theoretical insights and practical design guidelines for realizing stable, long-range SDC systems.
\end{abstract}

\section{Introduction}\label{sec:intro}
\begin{figure}[t]
	\centering
	\includegraphics[width=3.2in]{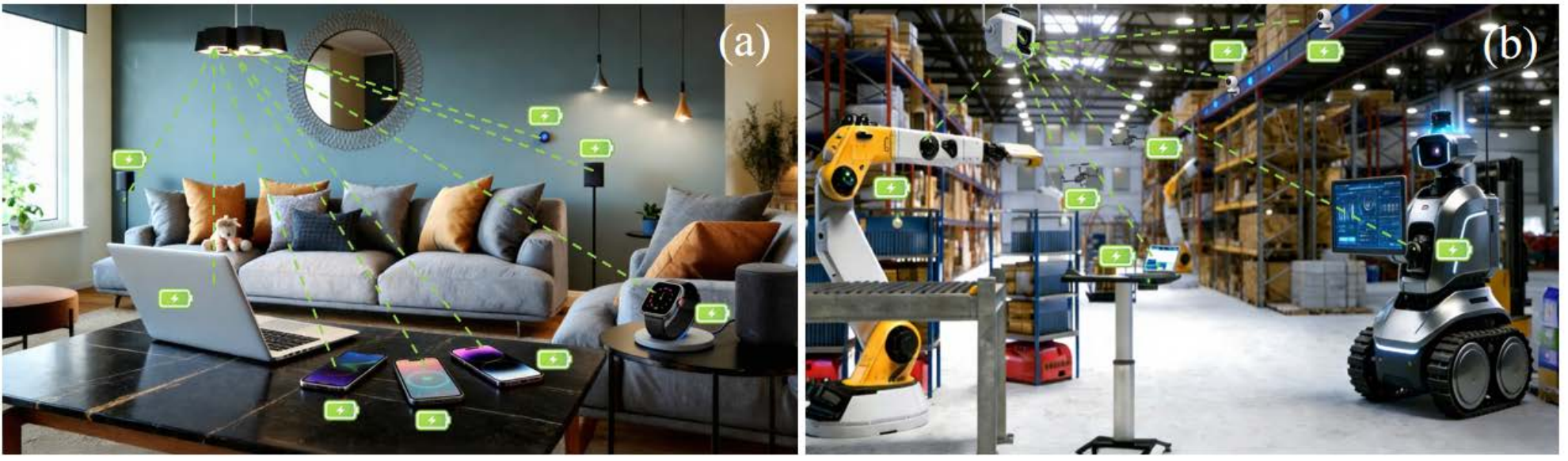}
	\caption{ \mll SDC application scenario diagram (a) indoor smart living (b) industrial intelligence.\mrr}
	\label{fig:scen}
    \vspace{-15pt}
\end{figure}

\IEEEPARstart{W}{ith} {\mll the pervasive growth of Internet of Things (IoT) devices such as mobile phones, smart speakers, electronic watches, and unmanned aerial vehicles (UAVs), the efficient delivery of both power and data has become increasingly critical~\cite{WPT_review1, WPT_review2, WPT_review3, WOC_review1,Krishnamoorthy2025-jf,Weng2024-pw,Zhang2024-kx,10271530}.\mrr} Simultaneous light information and power transfer (SLIPT) is a promising technology to address these demands, leveraging light's inherent advantages including high power density, large bandwidth, and immunity to electromagnetic interference (EMI)~\cite{SLIPT_concept1,laser_SLIPT_advantage,laser_SLIPT_challenge1}. Spatially distributed cavity (SDC)-based laser systems have emerged as a feasible solution for improving laser alignment mobility and safety in a variety of applications, including wireless power transfer (WPT), optical wireless communications (OWC), and SLIPT. Since the carrier is an intracavity laser that resonates within the distributed cavity structure, these systems, which include resonant beam charging (RBC), resonant beam communications (RBCom), and resonant beam SLIPT (RB-SLIPT), are also known as resonant beam systems (RBS). \mll As illustrated in Fig.~\ref{fig:scen}, the application scenario diagram of the SDC demonstrates its capability to provide power supply and data communication for indoor intelligent mobile terminals\mrr. Furthermore, RBS has also been explored for high-accuracy position and 3D pose estimation applications~\cite{Liu2025-yr}.

\begin{figure}[t]
	\centering
	\includegraphics[width=3.2in]{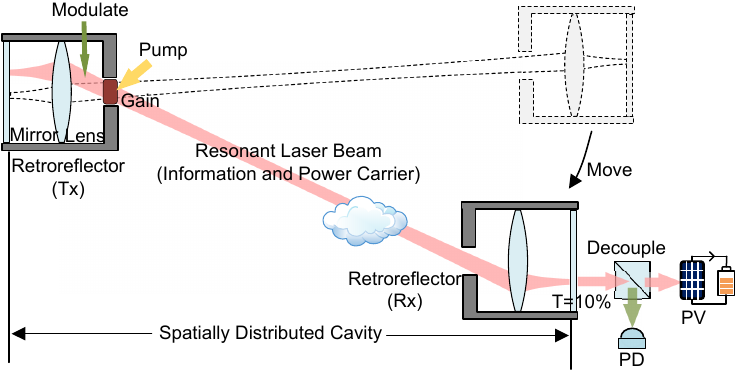}
	\caption{Spatially distributed cavity laser for mobility enhanced SLIPT. \mll(Tx: transmitter; Rx: receiver; PV: photovoltaic cells; PD: photodiode)\mrr}
	\label{fig:scenario}
    \vspace{-15pt}
\end{figure}

As depicted in Fig.~\ref{fig:scenario}, a fundamental SDC-based SLIPT system configuration typically employs two retroreflectors positioned at the transmitter~(Tx) and the receiver~(Rx), respectively, forming a laser resonator. An gain medium (laser crystal), pumped by a suitable source, is placed within the transmitter to provide optical power gain for the oscillating light inside the cavity. Partial of the resonant beam is released by the output mirror; and then, the information and the power is decoupled to the photon detector (PD) and the photovoltaic cells (PV), respectively. A key advantage of SDCs is their self-alignment characteristic, facilitated by the retroreflective property at both ends. 

{\mll By obviating the need for sophisticated acquisition, pointing, and tracking (APT) modules that are indispensable to traditional laser systems, this self-alignment feature significantly enhances system mobility, thereby rendering it highly suitable for mobile IoT devices such as UAVs and autonomous mobile robots (AMRs) that demand dynamic locomotion during charging or communication processes. Moreover, given that the resonant beam is confined within the cavity, any obstruction or foreign object intruding into the beam path will immediately disrupt the oscillation, thereby forming an intrinsic safety mechanism. This mechanism is not only particularly crucial for high-power transmission scenarios but also essential for IoT deployments in smart cities or industrial environments where high-power operation is mandated to be hazard-free. The receiver, which integrates a compact cat’s-eye retroreflector and photovoltaic devices, can be embedded into wireless sensors, drones, and other mobile IoT devices that are constrained by limited space for bulky batteries or dedicated charging interfaces. This feature enables such devices to operate continuously without the need for frequent battery replacement or wired charging, which is particularly valuable for high-power and electromagnetically sensitive applications.\mrr}

The concept of SDC dates back to 1974, when G. J. Linford proposed a system with two corner cube retroreflectors to support very long laser cavities for environmental sensing~\cite{a190318.02}. In 1980, Linford further designed a secure laser communication system utilizing SDC to prevent unintended detection~\cite{US4209689}. More recently, in 2016, Wi-Charge Company disclosed a patent describing a distributed coupled resonator laser based on two telecentric cat's eye retroreflectors (CER)~\cite{WO2016125155A1}. {\mll The same year, Liu \emph{et al.} presented comprehensive investigations into SDC-based WPT systems, wherein the technology’s salient advantages were elucidated, including mobility, intrinsic safety, multiple-receiver charging capability, high power delivery, compact size, and EMI-free operation. Collectively, these features were shown to effectively meet the core technical requirements of IoT application scenarios.~\cite{Liu-2016}\mrr}. Subsequently, in 2018, Q. Zhang \emph{et al}. developed an analytical model for a distributed laser charging (DLC) system employing two corner cube retroreflectors~\cite{Zhang2018-sg}.  M. Xiong \emph{et al}. proposed an RBCom system using SDC to mitigate challenges related to beam attenuation and tracking~\cite{Xiong2019-qb, Xiong2022-vh}. Addressing the need for multiple receivers, J. Lim \emph{et al}. proposed a wireless optical power transfer system that utilized spatial wavelength division and distributed laser cavity resonance, achieving $1.7$~mW over 1~m transmission distance with optical gratings and corner cubes~\cite{Lim2019-bs}. In the same year, W. Wang \emph{et al}. demonstrated an RBC system based on a quasi-SDC structure (dual-mirror cavity with spatially separated deployment), experimentally achieving $2$-W charging power over $2.6$~m. However, this specific implementation, based on a dual-mirror cavity, lacked the inherent mobility of typical SDC designs due to its alignment requirements~\cite{Wang2019-ko}.

Significant advancements continued in subsequent years. In 2021, W. Wang \emph{et al}. experimentally validated a Nd:YVO$_4$ thin disk SDC laser employing two CERs, demonstrating an improved alignment-free range of $-13^{\circ}$ to $13^{\circ}$~\cite{Wang2021-oo}. Further enhancing performance, M. Xiong \emph{et al}. proposed a low-diffraction-loss SDC design based on focusing CERs, which provide both retroreflection and focusing capabilities, and established an analytical model for output power calculation~\cite{Xiong2021-xs}. For simulating the intracavity light field, M. Liu \emph{et al}. introduced a Fox-Li-based algorithm to compute light field distribution and diffraction loss in SDCs~\cite{Liu2021-po}.

Since 2022, experimental research on SDC-based systems has become more prevalent and practical. Q. Liu \emph{et al}. developed a complete RBC system utilizing focusing CERs and a thin disk Nd:YVO$_4$ crystal, demonstrating mobile charging of a smartphone over $2$~m with received optical power up to $5$~W, resulting in $0.6$~W of electrical charging power~\cite{Liu2022-jc}. N.~Javed \emph{et al}. presented a WPT system incorporating an erbium-doped fiber amplifier~(EDFA) and ball lens retroreflectors~\cite{Javed2022-fl}. Q. Sheng \emph{et al}. explored the use of Nd:GdVO$_4$ crystal in an SDC and successfully enhanced the receiver's field of view (FoV) by compensating the field-curvature~\cite{Sheng2022-wr}. Z. Zhang \emph{et~al}. demonstrated $86.3$-mW optical power delivery over $200$-cm using vertical external-cavity surface-emitting lasers (VECSELs) as the gain medium, paving the way for semiconductor-based SDCs with potential for improved electro-optical efficiency~\cite{Zhang2022-tz}. In 2023, Q. Sheng \emph{et al}. reported an SDC system achieving $5$-W output optical power over $1$ to $5$~m with a receiving CER FoV up to $\pm 30^{\circ}$~\cite{Sheng2023-tf}. They also proposed a telescope-based SDC design in the same year, demonstrating $1.3$-W electrical power output at the receiver over a $5$-m transmission distance~\cite{Sheng2023-rj}. Investigating receiver's FOV, Y. Zuo \emph{et al}. designed a ball-lens CER, showing rotation tolerances up to $38^{\circ}$ in an SDC configuration. Furthermore, two-coupled SDC designs have been investigated in \cite{Sheng2023-rj,Deng2024-rc} to potentially mitigate safety risks and reduce transmission loss over extended distances.

\begin{figure*}[t]
	\centering
	\includegraphics[width=5.5in]{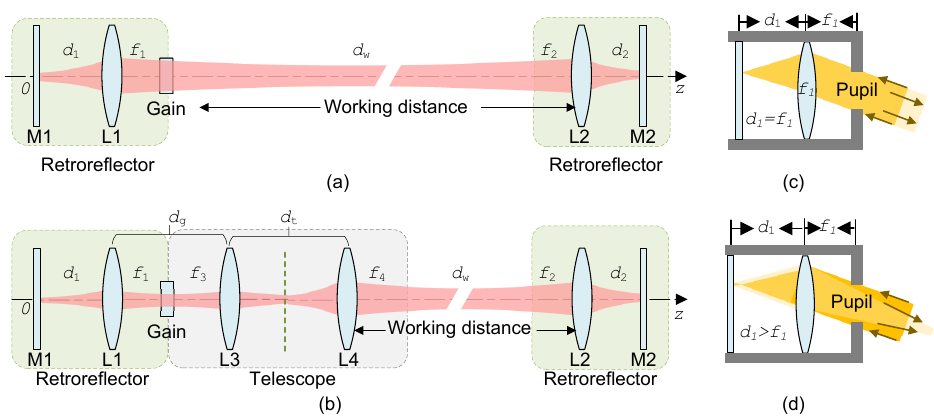}
	\caption{\mll System design of the duplex simultaneous light information and power transfer system based on a coupled spatially distributed cavity: (a) fundamental deployment, (b) cavity with internal telescope,  (c) common telecentric cat's eye retroreflector, (d) retroreflector with focusing ability. M1, M2: mirrors; L1--L4: lenses; $d1$, $d2$, $f1$--$f4$, $d_w$: intervals between elements; $f1$--$f_4$ are also focal lenses of L1--L4, respectively\mrr}
	\label{fig:sysdesign}
\end{figure*}

Theoretical investigations have also developed significantly. Studies have analyzed safety aspects related to foreign object intrusion into the optical cavity~\cite{Fang2021-ac} and proposed simulation methods for analyzing stable light intensity distribution within SDCs~\cite{Liu2021-po}. Optimization algorithms have been developed for asymmetric SDC designs~\cite{Xiong2022-vk}, and analytical models were established for optimizing the FoV of SDC systems~\cite{Han2024-ts}. While these studies cover various aspects, existing research, particularly experimental demonstrations, primarily focuses on meter-level transmission~(working) distances. There remains a notable lack of theoretical guidance specifically addressing the fundamental limitations of SDC technology for stable resonant beam formation over longer transmission distance. Unlike traditional external-cavity lasers, the spatially separated nature of SDCs means that the stability of the resonator imposes strict limits on the maximum achievable cavity length. Although stable resonators can be theoretically designed under ideal conditions, practical manufacturing tolerances and alignment errors inevitably introduce deviations from the optimal operational point, consequently limiting the achievable cavity length and thus the maximum transmission distance. Moreover, the trade-off between the cross-sectional area of the gain medium and the magnification of any integrated telescope optics represents another critical, yet underexplored, determinant factor in long-range resonant beam formation.

Motivated by these challenges and the prospect of extending SDC applications to longer ranges, this work conducts a rigorous theoretical and analytical investigation. The main contributions of this paper are summarized as follows:
\begin{itemize}
	\item We establish a comprehensive theoretical framework for analyzing the stability of telescope-enhanced spatially distributed cavity (SDC) laser systems. This framework identifies critical parameters influencing resonator stability and defines the boundaries for stable resonant beam formation across varying operational distances.
	\item We propose analytical methods, including a binary-search-based Monte Carlo simulation and a linear approximation technique, to quantify the maximum allowable manufacturing and assembly tolerances. These methods provide practical design guidelines crucial for achieving stable SDC operation at extended ranges.
	\item We identify several significant factors requiring careful consideration in the design of long-range SDC systems, including a detailed investigation into the relationship between the gain medium cross-sectional area and the magnification of the telescope.
	\item \mll We established an experimental platform to validate the performance of the  telescope-enhanced SDC system. \mrr We provide practical design and assembly guidelines for developing long-range SDC systems, specifically addressing parameter selection and adjustment procedures.
\end{itemize}

The remainder of this paper is structured as follows: Section~\ref{sec:design} describes the system structure of a SDC, detailing both a basic SDC configuration and a telescope-based design. Section~\ref{sec:model} presents the analytical system model and introduces the proposed algorithms for tolerance analysis. Numerical results and detailed discussions are provided in Section~\ref{sec:result}. Finally, Section~\ref{sec:con} concludes the paper.

\section{System Structure}
\label{sec:design}

This section details the architecture of the SDC, outlining its constituent components and the underlying mechanism enabling self-alignment and self-tracking. These features are critical for facilitating mobile WPT, OWC, and passive positioning. We first describe the basic SDC configuration, which utilizes two retroreflectors, and then introduce an enhanced version incorporating a telescope to potentially extend the transmission range or increase the field of view.

As illustrated in Fig.~\ref{fig:sysdesign}~(a), the fundamental SDC architecture comprises two retroreflectors, each consisting of a lens-mirror pair (M1-L1 and M2-L2), and a gain medium. Typically, in a SDC-based WPT or OWC system, one retroreflector is situated at the transmitter and the other at the receiver. The gain medium is integrated within the transmitter, precisely positioned in the pupil of the retroreflector. As depicted in Fig.~\ref{fig:sysdesign}~(c) and (d), the pupil is defined as an aperture located at the focal point of the lens, which is essential for the retroreflection property. A ray passing through the pupil will be reflected directly back towards its source. The distance between the lens and the mirror, denoted as $d_1$, governs the focusing behavior. When $d_1$ equals the lens focal length $f_1$, the retroreflector exhibits a non-focusing characteristic. Conversely, if $d_1 > f_1$, the retroreflector can actively focus the incident beam. These behaviors will be theoretically validated in the subsequent section. Due to the retroreflection occurring at both the transmitter and receiver ends, photons generated within the pumped gain medium are effectively confined and oscillate within this dual-retroreflector cavity.

The gain media utilized in SDC systems are typically crystals; however, alternative materials such as glasses, organic compounds, and semiconductors are also viable. In this study, we utilize an Nd:YVO$_4$ crystal as a representative example~\cite{Liu2022-jc,Wang2021-oo,Zuo2023-fx}. This crystal absorbs pump light at $880$~nm and amplifies light at $1064$~nm, which is called in-band pumping. Consequently, the optical power circulating within the SDC is progressively increased as photons traverse the gain medium during each round trip.

The remarkable self-aligning capability of the system stems directly from the inherent retroreflective characteristic. An intuitive, though not entirely precise, explanation is that photons originating from the gain medium are consistently reflected back to the gain medium by both retroreflectors, forming the oscillation. However, a more rigorous understanding requires considering that retroreflectors inherently invert the input image: photons generated at the top of the gain medium, for instance, will arrive at the bottom after reflection. Therefore, a comprehensive analysis must consider the collective behavior of all photons within the cavity. Wave optics theory provides an efficient tool for studying beam propagation in such systems. Previous research has conclusively demonstrated the existence of an oscillating optical path between the two retroreflectors that exhibits exceptionally low diffraction loss~\cite{Liu2022-jc,Liu2021-po}. A laser beam spontaneously forms along this resonant path even when the retroreflectors are not perfectly aligned along the same optical axis, thereby enabling the mobility of both the transmitter and receiver ends.

As illustrated in Fig.~\ref{fig:sysdesign}~(b), the enhanced SDC incorporates a telescope within the transmitter. This telescope, composed of two lenses (L3 and L4), serves to either compress the beam spot size at the gain medium or enlarge the effective field of view of the transmitter. Under the assumption that a plane mirror (represented by the green dashed line in Fig.~\ref{fig:sysdesign}) is hypothetically placed at the beam waist between lenses L3 and L4, the combination of the telescope and the retroreflectors can be conceptually viewed as two coupled SDCs. The left portion of the telescope (L3) paired with the first retroreflector (L1-M1) forms one effective SDC, while the right portion (L4) paired with the second retroreflector (L2-M2) forms another. This analogy clarifies why the inclusion of the telescope does not compromise the self-aligning property. This assumption holds because the wavefront curvature radius at the beam waist is infinite, effectively resembling a plane mirror. The telescope configured in this manner effectively possesses two pupils at both sides. It functions by compressing the light spot at its right-side pupil and projecting a corresponding image onto its left-side pupil. We will learn in the next section that a small light spot in the gain medium is required to improve the output power.

In the subsequent sections, we will quantitatively investigate the influence of various system parameters, including the distances between components and the focal lengths of the lenses. It is important to note that while distance parameters such as $d_1$, $d_2$, $d_{\rm g}$, $d_{\rm t}$, and $d_{\rm w}$ are adjustable, their acceptable ranges for stable operation vary significantly~\cite{a181218.01}. Specifically, the gain medium is nominally positioned at the focal image plane of both L1 and L3, implying that the ideal interval between L1 and L3, $d_{\rm g}$, should equal $f_1+f_3$. However, accounting for potential fabrication tolerances, we introduce a variable $\delta d_{\rm g}$ to represent the deviation from this ideal distance in our subsequent analysis. The interval between lenses L3 and L4, $d_{\rm t}$, should be close to $f_3+f_4$. This distance is adjustable and critical for finding a stable operating point for the SDC with the telescope. We will further explore the characteristics of $d_{\rm t}$'s adjustment range. The working distance, $d_{\rm w}$, is defined as the separation between the transmitter's pupil and the receiver's pupil. The total cavity length, represented by the interval between L4 and L2 (for the version with a telescope) or L1 and L2 (for the basic version), is thus given by $f_4+d_{\rm w}+f_2$ or $f_1+d_{\rm w}+f_2$, respectively. In the next section, we will establish a comprehensive system model to analyze the stability region of this resonator configuration under variations in these distance parameters and determine their boundaries for stable operation. Furthermore, we will investigate the principles guiding parameter selection and examine the tolerance levels acceptable during fabrication.

\section{Theory and Algorithm}
\label{sec:model}

Optical resonators can be effectively analyzed using the matrix method, also known as the ABCD matrix or ray transfer matrix method~\cite{a181221.01}. In this approach, different optical components such as plane mirrors, concave mirrors, convex lenses, and free-space propagation are represented by specific $2 \times 2$ matrices. The transformation of a ray's position and angle as it passes through the system is calculated by multiplying these optical matrices. Crucially, the overall system matrix is obtained by multiplying the individual component matrices in the order \emph{opposite} to the beam propagation direction.

\subsection{Stable Region Width}
\label{sec:stable_region}

For a plane mirror, the ray transfer matrix $\bm{M}_{\rm M}$ is the identity matrix, representing no change in ray position or angle upon reflection (considering propagation along the optical axis). A common retroreflector that simply reverses the beam path without focusing is represented by the matrix $-\bm{M}_{\rm M}$. This matrix effectively reverses both the ray's angle and its displacement from the optical axis. Using these fundamental matrices, the matrix for a specific focusing retroreflector system, consisting of a mirror between two free-space sections and lenses, is deduced as~\cite{Xiong2022-xj}:
\begin{equation}
	\begin{aligned}
		\bm{M}_{\rm RR}&=\bm{M}_{\rm T}(f)\bm{M}_{\rm L}(f)\bm{M}_{\rm T}(d)\bm{M}_{\rm M}\bm{M}_{\rm T}(d)\bm{M}_{\rm L}(f)\bm{M}_{\rm T}(f)\\
		&=
		\begin{bmatrix}
			1 & 0 \\
			-1/f_{\rm eqL} & 1
		\end{bmatrix}
		\begin{bmatrix}
			-1 & 0 \\
			0 & -1
		\end{bmatrix}
		\begin{bmatrix}
			1 & 0 \\
			-1/f_{\rm eqL} & 1
		\end{bmatrix}
	\end{aligned},
\end{equation}
where $\bm{M}_{\rm L}(f)$ is the matrix of lens with focal length $f$, $\bm{M}_{\rm T}(d)$ represents a translation or drift of length $d$, $f_{\rm eqL}$ is the equivalent focal length of the focusing retroreflector, defined by:
\begin{equation}
	f_{\rm eqL}=\frac{f^2}{d-f}.
\end{equation}
From the structure of $\bm{M}_{\rm RR}$, it can be seen that this retroreflector configuration is equivalent to a common retroreflector (represented by the matrix $\begin{bmatrix}-1 & 0 \\ 0 & -1\end{bmatrix}$) combined with a lens of focal length $f_{\rm eqL}$. If $d>f$, $f_{\rm eqL}$ is positive, indicating the matrix exhibits the characteristic of a focusing retroreflector. Conversely, if $d<f$, $f_{\rm eqL}$ is negative, and the matrix performs as a defocusing retroreflector. By appropriately adjusting the equivalent focal length $f_{\rm eqL}$ (which depends on $f$ and $d$), the SDC can be configured to operate in a stable state, minimizing diffraction losses.

{\mll For a specific telescope configuration utilized within this SDC system, characterized by lens focal lengths $f_{\rm 3}$ and $f_{\rm 4}$, and an interval $d_{\rm t}$ between the lenses, the ray transfer matrix describing propagation from one pupil plane to another is derived as:
\begin{equation}
	\begin{aligned}
		\bm{M}_{\rm TS}&=\bm{M}_{\rm T}(f_{\rm 4})\bm{M}_{\rm L}(f_{\rm 4})\bm{M}_{\rm T}(d_{\rm t})\bm{M}_{\rm L}(f_{\rm 3})\bm{M}_{\rm T}(f_{\rm 3})\\
		&=
		\begin{bmatrix}
			-M_{\rm tel} & 0 \\
			-\dfrac{f_{\rm 3}  + f_{\rm 4}- d_{\rm t}}{f_{\rm 3}f_{\rm 4}} & -\dfrac{1}{M_{\rm tel}}
		\end{bmatrix}
	\end{aligned},
	\label{equ:teles}
\end{equation}
where $M_{\rm tel} = {f_{\rm 4}}/{f_{\rm 3}}$ represents the angular magnification of the telescope.

For a telescope-based SDC, the round-trip transfer matrix, incorporating the telescope module, is expressed as:
\begin{equation}
	\begin{aligned}
		\bm{M}_{\rm SDC,t} =&
		\bm{M}_{\rm T}(d_{\rm 1}) \bm{M}_{\rm L}(f_1) \bm{M}_{\rm T}(d_{\rm g}) \bm{M}_{\rm L}(f_3)\\
		&\bm{M}_{\rm T}(d_{\rm t})\bm{M}_{\rm L}(f_4)\bm{M}_{\rm T}(f_4) \bm{M}_{\rm T}(d_{\rm w}) \bm{M}_{\rm T}(f_2)\\
		& \bm{M}_{\rm L}(f_2) \bm{M}_{\rm T}(d_2)\bm{M}_{\rm M}\bm{M}_{\rm T}(d_2)\bm{M}_{\rm L}(f_2)\bm{M}_{\rm T}(f_2)\\
		&\bm{M}_{\rm T}(d_{\rm w})\bm{M}_{\rm T}(f_4)\bm{M}_{\rm L}(f_4)\bm{M}_{\rm T}(d_{\rm t})\bm{M}_{\rm L}(f_3)\\
		&\bm{M}_{\rm T}(d_{\rm g})\bm{M}_{\rm L}(f_1)\bm{M}_{\rm T}(d_{\rm 1})\bm{M}_{\rm M}.
	\end{aligned}
	\label{equ:MSDCt}
\end{equation}
Here, $d_1, d_2, f_1, f_2, d_{\rm w}$ are as defined for the general SDC, while $f_3$ and $f_4$ are the focal lengths of the lenses in the telescope module, $d_{\rm t}$ is the distance between $f_3$ and $f_4$, and $d_{\rm g}$ is the distance between the first lens L1 and the telescope module's first lens L3. The ideal design distance $d_{\rm g}$ is set as $d_{\rm g}=f_1+f_3$. If there is a manufacturing or assembly error $\delta_{\rm g}$, the actual distance becomes $d_{\rm g}=f_1+f_3+\delta d_{\rm g}$. We will analyze the impact of this error on system stability.\mrr}

Next, we deduce the fundamental mode (TEM$_{00}$) radius, which characterizes the spatial extent of the lowest-order Gaussian beam within the resonator. For a Gaussian beam, the complex number $q$ encapsulates both the wavefront radius of curvature $\rho$ and the TEM$_{00}$ mode radius $w_{\rm 00}$; that is~\cite{a181221.01}:
\begin{equation}
	\frac{1}{q} = \frac{1}{\rho} - j \frac{\lambda}{\pi w_{\rm 00}^2},
	\label{equ:1mq}
\end{equation}
where $\lambda$ is the wavelength of the stimulated laser and $j$ is the imaginary unit ($j^2 = -1$).

From \eqref{equ:1mq}, \mll we can see that $w_{00}$ of a Gaussian beam at a given location can be obtained from the $q$-parameter by extracting the imaginary part of its reciprocal~\cite{Xiong2021-xs}\mrr:
\begin{align}
	w_{00}=\sqrt{\frac{-\lambda}{\pi \Im[1/q]}},
	\label{equ:w00q}
\end{align}
where $\Im[\cdot]$ denotes the operator for extracting the imaginary part of a complex number. For a physically meaningful beam radius, $\Im[1/q]$ must be negative.

\mll If a Gaussian beam with an initial complex beam parameter $q_{\rm in}$ enters an optical system with a ray transfer matrix $\begin{bmatrix}A & B \\ C & D\end{bmatrix}$, the transformed complex beam parameter $q_{\rm out}$ at the output plane is obtained as~\cite{a181221.01}:\mrr
\begin{equation}
	q_{\rm out} = \frac{Aq_{\rm in} + B}{Cq_{\rm in} + D}.
	\label{equ:qout}
\end{equation}
By applying the ABCD law, we can compute the transformed $q$-parameter of the beam after it completes a full round trip within the resonator.

Let $\begin{bmatrix}A_0& B_0\\ C_0& D_0\end{bmatrix}$ be the ray-transfer matrix representing one complete round trip within the SDC, starting and ending at mirror M1. For a stable optical resonator, according to the self-consistent mode theory, the $q$-parameter of the circulating mode at any given plane must remain unchanged after one round trip. This means the light field profile is reproduced after each round trip at the same location. Thus, for the $q$-parameter $q_0$ at M1, we must have $q_0 = q_{\rm out}$ after one round trip starting with $q_{\rm in} = q_0$. \mll This self-consistency condition yields the following equations~\cite{a181221.01}: \mrr
\begin{equation}
	\left\{
	\begin{aligned}
		&q_{\rm 0} = \frac{A_{\rm 0}q_{\rm 0} + B_{\rm 0}}{C_{\rm 0}q_{\rm 0} + D_{\rm 0}},\\
		&A_{\rm 0}D_{\rm 0}-B_{\rm 0}C_{\rm 0}=1.
	\end{aligned}
	\right.
	\label{equ:qm1}
\end{equation}
The optical transfer matrix satisfies $A_0 D_0 - B_0 C_0 = 1$ because its unit determinant ensures the system's reversibility, allowing light rays in vacuum to trace back from output to input. 

Solving the quadratic equation for $q_0$ from the first line of \eqref{equ:qm1}, we obtain two possible solutions. However, to ensure the mode radius $w_{00}$ is a real value, the imaginary part of $1/q_0$ in \eqref{equ:w00q} must be a negative number. \mll This condition selects the physically acceptable solution for $1/q_0$~\cite{a181221.01}:\mrr
\begin{equation}
	\frac{1}{q_0}=-\frac{A_0-D_0}{2B_0}-\frac{j}{2|B|}\sqrt{4-(A_0+D_0)^2}.
\end{equation}
Then, the fundamental mode radius at $z=0$ (the location of M1) can be deduced from \eqref{equ:w00q} by substituting the derived $q_0$:
\begin{equation}
	w_{00}(z=0) = \sqrt{\frac{2 \lambda |B_{\rm 0}|}{\pi \sqrt{4 - (A_{\rm 0} + D_{\rm 0})^2}}}.
	\label{equ:w00sqr}
\end{equation}
From equation \eqref{equ:w00sqr}, we can see that a real solution for the mode radius exists only when the term under the square root in the denominator is positive, i.e., $4-(A_0+D_0)^2>0$. This inequality forms the basis of the resonator stability criterion. Generally, a stability parameter $g$ is defined for simplification, relating to the trace of the round-trip matrix:
\begin{equation}
	g=\frac{A_0 + D_0}{2}.
\end{equation}
Using this parameter, the formula for the fundamental mode radius at M1 can be rewritten in a simplified form:
\begin{equation}
	w_{00}(z=0)= \sqrt{\frac{\lambda |B_0|}{\pi \sqrt{1 - g^2}}}.
\end{equation}
For the mode radius to be a real value, the term $1-g^2$ must be positive, which leads to the following condition~\cite{a181224.01}:
\begin{equation}
	-1 <g < 1.
	\label{equ:stableC}
\end{equation}
\mll The formula~\eqref{equ:stableC} is the most important equation for this study. It's the rule that determines whether the laser system can produce a stable, usable beam.\mrr 

Based on the formula derivations above, it is clear that both the distance parameters $\bm{d}=(d_1,d_2,d_{\rm g},d_{\rm t},d_{\rm w})$ and the focal length parameters $\bm{f}=(f_1,f_2,f_3,f_4)$ influence the stability of the SDC. Typically, the focal lengths $\bm{f}$ are fixed by design and component selection, while the distance parameters $\bm{d}$ can be adjusted during assembly. A critical question in the design and implementation of such a system is to determine the acceptable range for an adjustable parameter while all other parameters are fixed, such that the SDC remains in a stable state. For instance, if all parameters except $d_{\rm t}$ are fixed, we can calculate the stable region width for $d_{\rm t}$. This width, denoted as $\mathcal{D}_{d_{\rm t}}$, is the difference between the values of $d_{\rm t}$ at the boundaries of the stable region where $g=1$ and $g=-1$:
\begin{equation}
	\mathcal{D}_{d_{\rm t}} = d_{\rm t}|_{g=1}-d_{\rm t}|_{g=-1}.
\end{equation}
\mll This equation tells us the range of $d_{\rm t}$ that will still allow for a stable cavity. It helps designers determine how much tolerance they have for manufacturing and assembly errors. \mrr 
Similarly, we can obtain the stable region width $\mathcal{D}_{d_1}$ for the parameter $d_1$. Calculating these boundary values involves finding the roots of the polynomial equations $g(d_i) = \pm 1$. This calculation can be performed using numerical equation-solving functions available in software like Matlab.

\subsection{Beam Radius Evolution}
\label{sec:beam_radius}

Following the determination of the resonator's stability and the fundamental mode properties at M1, the next step is to calculate the fundamental mode radius $w_{00}(z)$ at an arbitrary location $z$ along the beam path within the cavity. As established earlier, the mode radius at any point can be calculated from the complex beam parameter $q(z)$ using formula~\eqref{equ:w00q}. Therefore, the primary objective is to obtain $q(z)$. Having already obtained the self-consistent $q_0$ at M1 (which we define as $z=0$), we can calculate $q(z)$ by applying the ABCD law \eqref{equ:qout} using $q_0$ as the input parameter $q_{\rm in}$ and the ray-transfer matrix of the optical system situated between $z=0$ and the location $z$.

Let $\bm{M}_{\rm t}(z) = \begin{bmatrix}A_{\rm t} & B_{\rm t} \\ C_{\rm t} & D_{\rm t}\end{bmatrix}$ be the ray transfer matrix describing the propagation from the plane $z=0$ (M1) to the arbitrary plane $z$. This system matrix can be expressed as the following piecewise function, accounting for the sequence of optical elements:
\begin{equation}
	\bm{M}_{\rm t}(z) =\left\{
	\begin{aligned}
		&\bm{M}_{\rm T}(z - z_k) ,\quad\text{for } k=1\\
		&\bm{M}_{\rm T}(z - z_k) \prod_{i=1}^{k-1} \left[ \bm{M}_{\rm L}(\mathtt{f}_i) \bm{M}_{\rm T}(z_i - z_{i-1}) \right],\\ &~~~~~~~~~~~~~~~~~~~\text{for } k\ge 2 \text{ and } z_k < z \leq z_{k+1}\\
	\end{aligned}\right.
\end{equation}
where $z_k$ represents the cumulative distance from the starting plane (M1) to the end of the $k$-th complete segment of the optical system, and $\mathtt{f}_i~(i=1,2,\dots,k)$ is the focal length of the lens within the $i$-th segment. In this formulation, the SDC is conceptually divided into several segments, each typically containing a free-space propagation section followed by a lens or other optical element. The multiplicative operation denotes the ordered product of the matrices for all completed segments the beam passes through before reaching $z$. The final segment, within which the point $z$ is located, is not complete, and its free-space propagation matrix $\bm{M}_{\rm T}(z - z_k)$ is written on the left side of the product, consistent with the reverse-order multiplication rule of matrix optics. The cumulative distance $z_k$ represents the total length covered by all segments completed before the beam reaches the $k$-th optical element. It is calculated recursively by:
\begin{equation}
	z_k=\left\{
	\begin{array}{ll}
		0,  &\mbox{for } k=1\\
		z_{k-1}+\mathtt{z}_{k-1}, & \mbox{for } k \geq 2
	\end{array}
	\right.
\end{equation}
where $\mathtt{z}_k$ represents the length of the $k$-th segment, defined as the interval between the optical elements (mirror-lens or lens-lens) bounding that segment in the direction of beam propagation starting from M1. For the telescope-based SDC configuration described by \eqref{equ:MSDCt}, the component intervals can be defined as:
\begin{equation}
	(\mathtt{z}_k)_{k=1}^5=(d_1,~d_{\rm g},~d_{\rm t},~f_4+d_{\rm w}+f_2,~d_2).
\end{equation}
Once the matrix $\bm{M}_{\rm t}(z)=\begin{bmatrix}A_{\rm t}&B_{\rm t}\\C_{\rm t}&D_{\rm t}\end{bmatrix}$ from $z=0$ to $z$ is determined, we can obtain the complex beam parameter $q(z)$ at location $z$ using the ABCD law \eqref{equ:qout} with $q_0$.
Subsequently, it is straightforward to calculate the TEM$_{00}$ mode radius at $z$, $w_{00}(z)$, by substituting $q(z)$ into formula~\eqref{equ:w00q}. It is important to note that $w_{00}(z)$ represents the radius of the fundamental Gaussian mode. The actual beam in the resonator may be a superposition of multiple transverse modes. The ratio of the overall beam radius to the fundamental mode radius at any plane $z$ is characterized by a constant value, which is defined as the beam propagation factor $M^2$. For a pure TEM$_{00}$ mode, $M^2=1$. For  multimode beams, $M^2>1$.

\subsection{Tolerance Analysis}
\label{sec:tolerance}

The performance and stability of a real SDC system depend on the precise values of the distance parameters $\bm{d}=(d_1,d_2,d_{\rm g},d_{\rm t},d_{\rm w})$ and focal length parameters $\bm{f}=(f_1,f_2,f_3,f_4)$. In practical manufacturing and assembly, there are inherent tolerances. Let $\tau_{\rm d}$ and  $\tau_{\rm f}$ represent the tolerance associated with distance parameters and lens focal lengths, respectively. Due to the existence of these tolerances, the actual physical parameters ($d_i'$ and $f_k'$) in a fabricated system may deviate from their nominal design values ($d_i$ and $f_k$). These deviations can potentially cause the real system to operate in an unstable region. To guarantee that the SDC operates within the stable region despite these manufacturing and assembly variations, the stability criterion $|g|<1$, as shown in~\eqref{equ:stableC}, must be satisfied for all possible combinations of parameters within their respective tolerance ranges:
\begin{equation}
	\left|g(d_1',d_2',d_{\rm g}',d_{\rm t}',d_{\rm w}',f_1',f_2',f_3',f_4')\right|<1,
\end{equation}
where the actual parameters $d_i'$ and $f_k'$ are allowed to vary within the intervals defined by:
\begin{equation}
	\begin{aligned}
		&d_i'\in[d_i-\tau_{\rm d}, d_i+\tau_{\rm d}], \quad \text{for } i \in \{1, 2, {\rm g}, {\rm t}, {\rm w}\},\\
		&f_k'\in[f_k-\tau_{\rm f}, f_k+\tau_{\rm f}], \quad \text{for } k \in \{1,2,3,4\}.
	\end{aligned}
\end{equation}
\mll Generally, the tolerance of the lenses $\tau_{\rm f}$ is specified by the manufacturer or determined by design requirements, denoted here as a fixed value $\tau_{\rm f}^*$. The objective then becomes determining the maximum acceptable tolerance for the distance parameters, $\tau_{\rm d}^{\rm max}$, under the given design parameters and the fixed lens tolerance $\tau_{\rm f}^*$, such that the stability criterion is always met. We propose a binary-search-based Monte Carlo (BMC) simulation method that uniquely integrates the efficiency of binary search with the statistical robustness of Monte Carlo simulation, for long-range SDC systems to overcome the limitations of conventional tolerance analysis methods in complex resonant beam systems.\mrr

The pseudocode for the BMC algorithm is depicted in Algorithm~\ref{alg:bmc}. The process begins by calculating the nominal stability parameter $g_0$ for the system with the ideal design parameters $\bm{d}$ and $\bm{f}$. Subsequently, an iterative loop is executed to perform a binary search for the maximum acceptable $\tau_{\rm d}$. Within each iteration of the binary search loop, an internal loop conducts a Monte Carlo simulation (MCS). The MCS involves generating $N$ sets of random deviations ($\delta d_{\rm i}$ and $\delta f_{\rm k}$) for the distance and focal length parameters, respectively. These deviations are drawn from uniform distributions $\mathcal{U}[-\tau_{\rm d}, \tau_{\rm d}]$ for distances and $\mathcal{U}[-\tau_{\rm f}^*, \tau_{\rm f}^*]$ for focal lengths, where $\tau_{\rm d}=(l+u)/2$ is the current test value in the binary search (with $l$ and $u$ being the lower and upper bounds of the search interval for $\tau_{\rm d}$). For each generated set of deviations, the actual parameters $d_i' = d_i + \delta d_i$ and $f_k' = f_k + \delta f_k$ are computed, the corresponding round-trip matrix $M_{\rm SDC,t}$ is determined, and the resulting stability parameter $g$ is calculated. Here $\bm{d}'$, $\delta\bm{d}$, $\bm{f}'$, and $\delta\bm{f}$ represent the combination of $d_i'$  $\delta d_i$, $f_k'$ and $\delta f_k$, respectively. The maximum absolute value of $g$ encountered across all $N$ samples in the current MCS run, denoted $g_{\text{max}}$, is tracked. After completing the MCS, the algorithm checks if $g_{\text{max}}$ is less than or equal to 1, indicating that all tested configurations within the current $\tau_{\rm d}$ range are stable. If $g_{\text{max}} \leq 1$, the current $\tau_{\rm d}$ is a possible maximum, so the lower bound $l$ of the binary search is updated to $\tau_{\rm d}$. Conversely, if $g_{\text{max}} > 1$, the current $\tau_{\rm d}$ is too large, and the upper boundary $u$ is reduced to $\tau_{\rm d}$. After a specified number of iterations, $I$, the value of $\tau_{\rm d}$ converges towards the maximum acceptable tolerance $\tau_{\rm d}^{\rm max}$.

\begin{algorithm}
	\caption{Binary-Search-Based Monte Carlo Simulation for Tolerance Analysis}
	\label{alg:bmc}
	\begin{algorithmic}[1]
		\State \textbf{Input:} Parameters $\bm{d}$ and $\bm{f}$, $\tau_f^*$, samples $N$, iterations $I$, $\tau_d$ bounds $[l, u]$
		\State \textbf{Output:} Max $\tau_d^{\text{max}}$
		\For{$iter = 1$ to $I$}
		\State $\tau_d \gets (l + u) / 2$
		\State $g_{\text{max}} \gets 0$
		\For{$n = 1$ to $N$}
		\State $\delta \bm{d} \sim \mathcal{U}[-\tau_d, \tau_d]$, $\delta \bm{f} \sim  \mathcal{U}[-\tau_f^*, \tau_f^*]$
		\State $\bm{d}' \gets \bm{d} + \delta \bm{d}$, $\bm{f}' \gets \bm{f} + \delta \bm{f}$
		\State $M_{\rm SDC,t} \gets \mathsf{sdc\_matrix}(\bm{d'}, \bm{f'})$
		\State $g \gets \mathsf{g\_parameter}(M_{\rm SDC,t})$
		\State $g_{\text{max}} \gets \mathsf{max}(g_{\text{max}}, |g|)$
		\EndFor
		\If{$g_{\text{max}} \leq 1$}
		\State $l \gets \tau_d$
		\Else
		\State $u \gets \tau_d$
		\EndIf
		\EndFor
		\State \Return $\tau_d^{\text{max}} \gets \tau_d$
	\end{algorithmic}
\end{algorithm}

However, the BMC simulation can be computationally expensive and slow, especially when a large number of samples, $N$, is required to achieve sufficient statistical confidence. Hence, we propose an alternative \emph{linear approximation} method to estimate $\tau_{\rm d}^{\rm max}$ more quickly. The actual stability parameter $g'$ for a system with tolerances $(\tau_{\rm d}, \tau_{\rm f})$ can be related to the nominal stability parameter $g(\bm{d},\bm{f})$ and a deviation term $\Delta g(\tau_{\rm d},\tau_{\rm f})$ induced by these tolerances. The absolute value of the perturbed stability parameter $|g'|$ must satisfy the inequality:
\begin{equation}
	\left|g'(\bm{d},\bm{f}, \tau_{\rm d},\tau_{\rm f})\right|\leq \left|g(\bm{d},\bm{f})\right|+|\Delta g(\tau_{\rm d},\tau_{\rm f})|.
\end{equation}
The goal is to find the maximum possible $\tau_{\rm d}^{\rm max}$ such that for any combination of deviations within the ranges $[-\tau_{\rm d}^{\rm max}, \tau_{\rm d}^{\rm max}]$ for distances and $[-\tau_{\rm f}^{*}, \tau_{\rm f}^{*}]$ for focal lengths, the resulting resonator is stable. Under the assumption of small deviations, the worst-case variation of $g$, $|\Delta g(\tau_{\rm d},\tau_{\rm f})|$, can be approximated using a linear Taylor expansion around the nominal design point. The total worst-case deviation is approximately the sum of the absolute contributions from each parameter's maximum deviation:
\begin{equation}
	|\Delta g(\tau_{\rm d},\tau_{\rm f})| \approx \sum_{i} \left| \frac{\partial g}{\partial d_i} \right| \tau_{\rm d} + \sum_{k} \left| \frac{\partial g}{\partial f_k} \right| \tau_{\rm f}.
\end{equation}
Here, the partial derivatives $\frac{\partial g}{\partial d_i}$ and $\frac{\partial g}{\partial f_k}$ represent the sensitivity of the stability parameter $g$ to changes in each distance and focal length parameter, respectively, evaluated at the nominal design values.
Considering the stable condition restriction $|g'| < 1$, we can approximate the boundary of the stable region in the presence of tolerances by:
\begin{equation}
	\left|g(\bm{d},\bm{f})\right|+\sum_{i} \left| \frac{\partial g}{\partial d_i} \right| \tau_{\rm d} + \sum_{k} \left| \frac{\partial g}{\partial f_k} \right| \tau_{\rm f} < 1.
\end{equation}
Since the focal length tolerance $\tau_{\rm f}^*$ is given and fixed, we can solve this inequality for $\tau_{\rm d}$ to find the maximum acceptable tolerance $\tau_{\rm d}^{\rm max}$ under this linear approximation:
\begin{equation}
	\tau_{\rm d}^{\rm max}=\dfrac{1-|g(\bm{d},\bm{f})|-S_{\rm f}\tau_{\rm f}^*}{S_{\rm d}},
\end{equation}
where
\begin{equation}
	\begin{aligned}
		S_{\rm d} &= \sum_{i} \left| \frac{\partial g}{\partial d_i} \right|, \\
		\quad S_{\rm f} &= \sum_{k} \left| \frac{\partial g}{\partial f_k} \right|.
	\end{aligned}
\end{equation}
In practice, the partial derivatives $\frac{\partial g}{\partial d_i}$ and $\frac{\partial g}{\partial f_k}$ can be estimated numerically by introducing a small perturbation to each parameter individually while holding others constant and observing the resulting change in $g$. All the contributions to the absolute variation of $g$ induced by these deviations are accumulated to model the worst-case condition.

\subsection{Output Power}
\label{sec:output_power}

The resonant beam circulates within the optical cavity, experiencing amplification from the gain medium and suffering various power losses. The output power extracted from a specific mirror (in this case, M2) is achieved at a steady state where the total optical gain within the cavity exactly balances the total cavity losses. Based on the classic Rigrod analysis for solid-state lasers, the output power can be approximated by the following equation~\cite{Xiong2021-xs}:
\begin{equation}
	P_{\text{out}} = \frac{\mathcal{T}_{\rm 2o} \pi a_{\rm g}^2 I_{\rm s}}{\left(1 + \sqrt{\frac{\mathcal{R}_2}{\mathcal{R}_1}}\right) \left(1 - \sqrt{\mathcal{R}_1 \mathcal{R}_2}\right)} \left[ \frac{l_{\rm g} \eta_{\rm c} P_{\rm in}}{I_{\rm s} V_{\rm g}} - \ln \frac{1}{\sqrt{\mathcal{R}_1 \mathcal{R}_2}} \right].
	\label{equ:rigrod}
\end{equation}
In this equation, $P_{\text{out}}$ is the extracted output power and should always be a non-negative number. $\mathcal{T}_{\rm 2o}$ represents the effective transmissivity from the gain medium interface towards the output side through mirror M2. $I_{\rm s}$ is the saturation intensity of the gain medium, a material property. $a_{\rm g}$ is the effective radius of the active region (pumped volume) within the gain medium, assuming a cylindrical geometry. $\mathcal{R}_1$ and $\mathcal{R}_2$ are the equivalent round-trip power reflectivities experienced by the resonant beam at the left side and right side of the gain medium, respectively, accounting for all losses and reflections encountered during a round trip to these interfaces. $l_{\rm g}$ is the physical thickness of the gain medium along the propagation axis. $\eta_{\rm c}$ is a combined efficiency of the pump power into the gain medium. $P_{\rm in}$ is the incident pump power. $V_{\rm g}$ is the volume of the active gain medium, calculated as $V_{\rm g}=\pi a_{\rm g}^2 l_{\rm g}$. The active area is the cross-sectional region of the gain medium that is effectively pumped and provides optical amplification to the circulating resonant beam.

The equivalent parameters $\mathcal{T}_{\rm 2o}$, $\mathcal{R}_1$, and $\mathcal{R}_2$ are computed by considering the series of optical elements and interfaces the beam encounters in a round trip: they are given by:
\begin{equation}
	\begin{aligned}
		\mathcal{T}_{\rm 2o}&=T_{\rm ARs,2o}T_{\rm air}T_{\rm M2},\\
		\mathcal{R}_{1}&=T_{\rm diff,1}T_{\rm ARs,1}R_{\rm M1},\\
		\mathcal{R}_{2}&=T_{\rm diff,2}T_{\rm ARs,2}T_{\rm air}R_{\rm M2}.
	\end{aligned}
	\label{equ:equivalent_parameters}
\end{equation}
Here, $T_{\rm ARs,2o}=T_{\rm ar}^7$ quantifies the total power transmission through all anti-reflection (AR) coatings traversed by the beam in the output path from the gain medium to the exterior of the cavity. $T_{\rm ar}=0.995$ is the specified power transmissivity of a single AR coating. Similarly, $T_{\rm ARs,1}=T_{\rm ar}^6$ and $T_{\rm ARs,2}=T_{\rm ar}^{14}$ represent the cumulative transmission through AR coatings on the path from the gain medium interface to M1 and M2, respectively, and back to the interface. $R_{\rm M1}$ and $R_{\rm M2}$ are the power reflectivities of mirrors M1 and M2, respectively. Typically, $R_{\rm M1}=0.999$ represents a high reflectivity (HR) coating, while $R_{\rm M2}<1$ acts as the output coupler, with its specific value depending on the desired output power coupling. $T_{\rm M2}=1-R_{\rm M2}$ is the power transmissivity of the output coupling mirror M2. $T_{\rm air}=\exp(-\alpha_{\rm air}d_{\rm w})$ is the attenuation of air transmission, where $\alpha_{\rm air}=10^{-4}$~m$^{-1}$ is the absorption coefficient of clear air. $T_{\rm diff}$ represents the power transmission factor associated with diffraction losses incurred by the resonant beam as it propagates through finite apertures of optical devices within the cavity. \mll We use the following empirical formula to estimate the diffraction loss factor occurring near the gain medium at both interfaces (accounted for in $\mathcal{R}_1$ and $\mathcal{R}_2$ via $T_{\rm diff,1}$ and $T_{\rm diff,2}$)~\cite{Xiong2021-xs}:\mrr
\begin{equation}
	T_{\rm diff}=1-\exp\left[-2\left(\dfrac{a_{\rm g}}{w_{00}(z_{\rm g})}\right)^2\right],
	\label{equ:diff}
\end{equation}
where $w_{00}(z_{\rm g})$ is the fundamental mode radius at the longitudinal position, $z_{\rm g}$, of the gain medium. The ratio $a_{\rm g}/w_{00}(z_{\rm g})$ is a measure of how well the beam is clipped by the gain medium's physical extent. Note that this formula provides an estimation and may not be strictly accurate for all cases, particularly for complex mode profiles or strong clipping. However, it is computationally efficient and provides a useful approximation for evaluating the impact of the gain medium aperture on beam transmission. This is particularly relevant because optical resonators are highly sensitive to intracavity losses, the diffraction loss calculated by this formula typically remains near zero, as expected for a well-aligned stable cavity.

\begin{figure*}[t]
	\centering
	\includegraphics[width=7in]{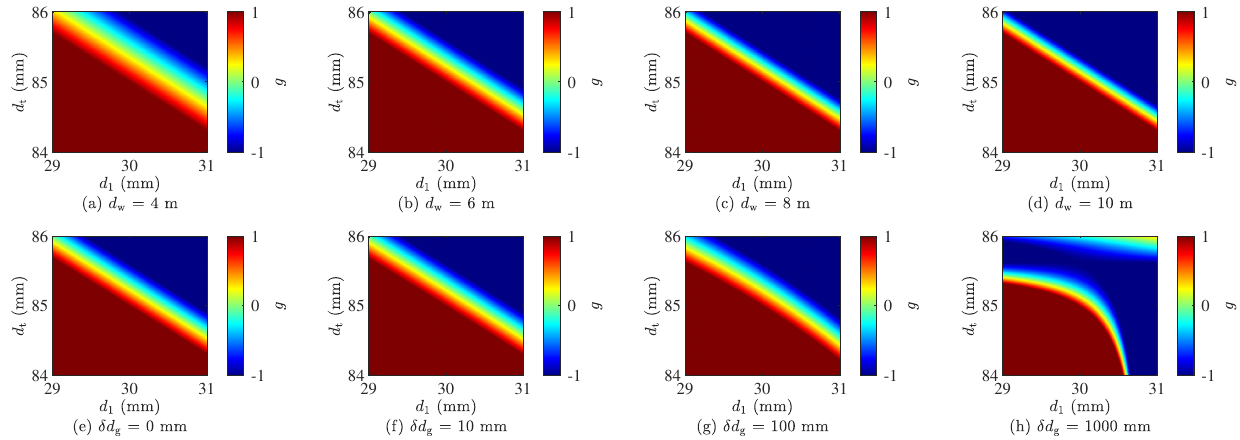}
	\caption{Stability parameter $g$ under different ($d_1,d_{\rm t}$): (a)--(d) shows different working distance $d_{\rm w}$, and (e)--(f) shows different deviation $\delta d_{\rm g}$.}
	\label{fig:G-D1DT}
\end{figure*}

\section{Results and Discussion}
\label{sec:result}

In this section, we investigate  key properties of the SDC system with emphasis on the characteristics of the stable operating region and the constraints on the working distance. Based on this analysis, we provide guidelines for system design and assembly. Since we have chosen Nd:YVO$_4$ as the gain medium, a material commonly used in practical resonant beam systems, the saturation intensity $I_{\rm s}=1.1976\times 10 ^ 7~\mbox{W}/\mbox{m}^2$. The radius of the active gain medium  is set to $a_{\rm g}=2.5$~mm, and its thickness is $l_{\rm g}=1$~mm. The combined pump coupling efficiency is $\eta_{\rm c}=0.439$~\cite{Xiong2022-kk}. The pump power is set to $P_{\rm in}=65$~W at a wavelength of $880$~nm, which stimulates a resonant beam at a wavelength of $1064$~nm. The reflectivity of the output coupling mirror M2 is set to $R_{\rm M2}=0.95$.

Unless otherwise specified, the analysis uses the following default distance parameters: $(d_1,d_2,d_{\rm g},d_{\rm t},d_{\rm w})=(30,30,55,85.4,6000)$~mm. The default focal length parameters are $(f_1,f_2,f_3,f_4)=(30,30,25,60)$~mm. In the subsequent analysis, parameters are either held at their default values, scaled by a stated factor relative to the basic configuration, or varied as independent variables as indicated.

\subsection{Stable Region Width}

\begin{figure}[t]
	\centering
	\includegraphics[width=3.2in]{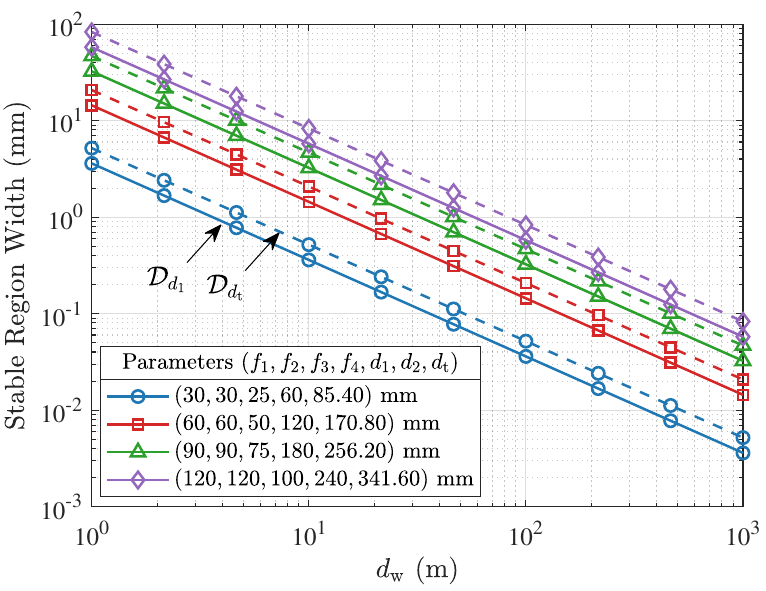}
	\caption{Stable region width under different working distance $d_{\rm w}$ (solid line: $\mathcal{D}_{d_1}$; dash line: $\mathcal{D}_{d_{\rm t}}$; $d_{\rm g}=f_1+f_3$).}
	\label{fig:Ddw}
\end{figure}

\begin{figure}[t]
	\centering
	\includegraphics[width=3.2in]{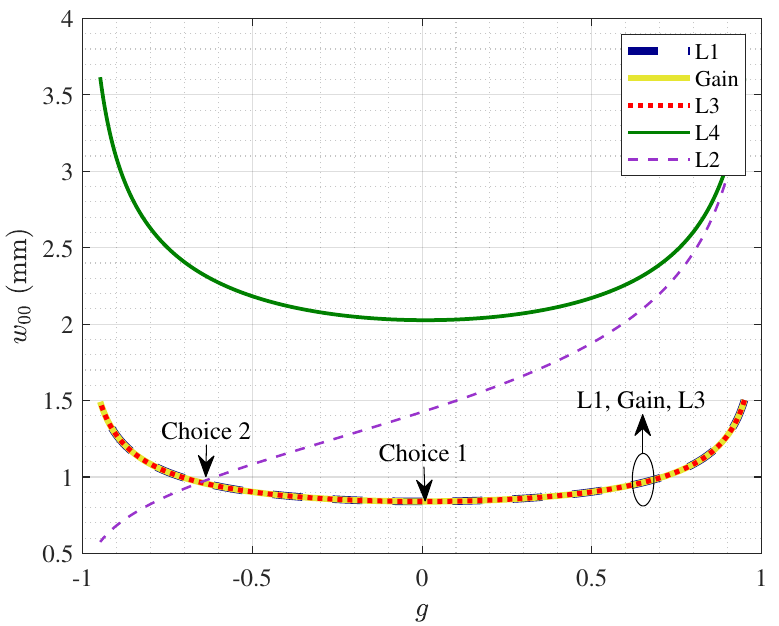}
	\caption{\mll TEM$_{00}$ mode radius at different optical devices varies with stable parameter $g$.\mrr }
	\label{fig:W00g}
\end{figure}

Figure~\ref{fig:G-D1DT} illustrates how the stability parameter $g$ varies with different values of $d_{\rm t}$ and $d_1$. The color coding represents different levels of the $g$ parameter. From Fig.~\ref{fig:G-D1DT} (a)--(d), we can observe that the width of the stable region in terms of $d_{\rm t}$ adjustment (i.e., $\mathcal{D}_{d_{\rm t}}$) remains relatively constant for varying $d_1$, provided $d_{\rm w}$ is fixed. However, as $d_{\rm w}$ increases, $\mathcal{D}_{d_{\rm t}}$ becomes narrower. This indicates that even if the parameter $d_1$ is not precisely manufactured or assembled in a real system, it is still possible to adjust $d_{\rm t}$ to achieve operation within the stable region. Nevertheless, achieving stable operation becomes more challenging and requires more rigorous adjustment precision for longer $d_{\rm w}$. Figure~\ref{fig:G-D1DT} (e) -- (h) show that small deviation in $d_{\rm g}$ ($\delta d_{\rm g}<10$~mm), have a negligible impact on stability. However, a very large deviation in $d_{\rm g}$ will distorts the pattern of the stable region. In practice,  $\delta d_{\rm g}$ is expected to be much smaller than $10$~mm, so this impact can typically be ignored. A notable characteristic observed in Fig.~\ref{fig:G-D1DT} is that the boundary $g=1$ remains constant despite significant changes in $d_{\rm w}$.

As depicted in Fig.~\ref{fig:Ddw}, the width of the stable region decreases as the working distance $d_{\rm w}$ increases. We consider four different configurations in this analysis. The basic configuration uses focal lengths $(f_1,f_2,f_3,f_4)=(30,30,25,60)$~mm, and the other three conditions scale these focal lengths by factors of 1, 2, and 3. We observe that at kilometer-level working distances, the stable region width is on the order of $0.001$~mm, which is extremely narrow. If the focal lengths of the lenses are increased, for instance, scaled by a factor of 3 times the basic configuration, the stable region width increases to the order of $0.01$~mm. \mll At this operational scale, the SDC system proposed in this work demonstrates theoretical feasibility for supporting kilometer-scale SLIPT applications\mrr. In all cases presented, the stable region width for $d_{\rm t}$ ($\mathcal{D}_{d_{\rm t}}$) is greater than that for $d_{\rm 1}$ ($\mathcal{D}_{d_{\rm 1}}$) at any given $d_{\rm w}$. This supports the conclusion that $d_{\rm t}$ related to the telescope structure is a more effective parameter for adjusting the operating point into the stable region.

\begin{figure}[t]
	\centering
	\includegraphics[width=3.2in]{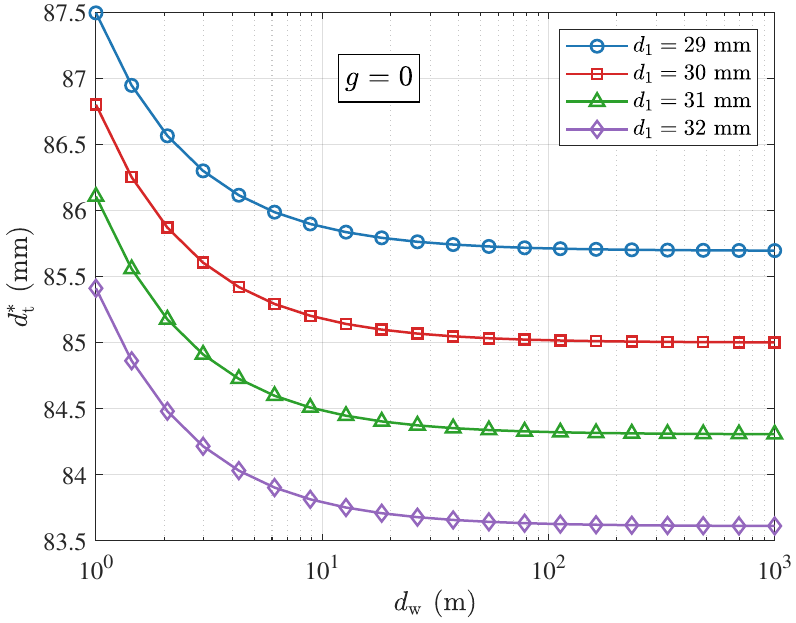}
	\caption{$d_{\rm t}^*$ (obtained by setting $g=0$) varies with working distance $d_{\rm w}$.}
	\label{fig:dtdw}
\end{figure}

\subsection{Relation Between $w_{00}$ and $g$ Parameters}
\label{sec:w00_g_relation}

Figure~\ref{fig:W00g} illustrates how the TEM$_{00}$ mode radii at various optical elements (L1, the gain medium, L3, L4, and L2) change as the stability parameter $g$ varies. We observe that the $w_{00}$ values at L1, the gain medium, and L3 are nearly identical and overlap as $g$ changes, indicating that the beam propagating between L1 and L3 is approximately collimated. The beam radius reaches its minimum value when $g=0$. This suggests that $g=0$ is an optimal design criterion because, as implied by equation \eqref{equ:diff}, a smaller mode radius at the gain medium location results in lower diffraction loss. From this figure, we also see that $w_{00}$ at L4 follows a similar trend to that at L3 as $g$ increases. In contrast, $w_{00}$ at L2 increases monotonically with $g$, a behavior different from the other locations. This characteristic should be considered in the design process, as a small beam spot at L2 is generally desired to minimize the size requirements for the receiver optics. Consequently, another possible design choice could be to set $g$ at or near the intersection point of the curves for $w_{00}$ at L2 and the gain medium, although this strategy might impose stricter requirements on component tolerances and assembly.

Based on the analysis presented above, we understand that $d_{\rm t}$ can serve as an effective adjustment parameter to enable the system to operate at $g=0$, which is an optimal operating point for minimizing the mode size at the gain medium. Hence, we next investigate the behavior of $d_{\rm t}$ under the condition $g=0$ (denoted as $d_{\rm t}^*$), as depicted in Fig.~\ref{fig:dtdw}. This figure shows that $d_{\rm t}^*$ decreases as the working distance $d_{\rm w}$ increases, and it appears to approach a fixed asymptotic value. When different values for $d_1$ are considered, we observe that this asymptotic line shifts upwards with decreasing $d_1$. This trend can be attributed to the gradual shrinking of the stable zone and the fixed $g=1$ boundary, a phenomenon also intuitively demonstrated in Fig.~\ref{fig:G-D1DT}.

\begin{figure}[t]
	\centering
	\includegraphics[width=3.2in]{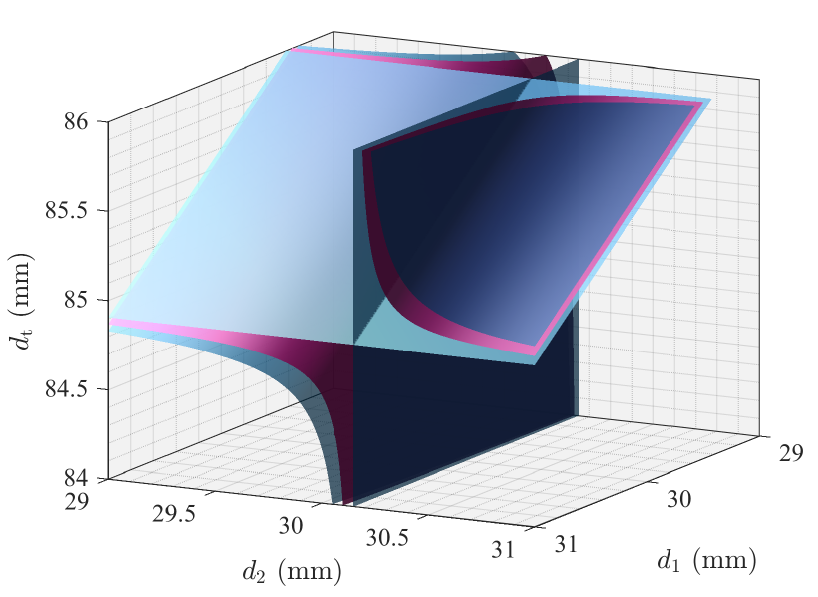}
	\caption{3D perspective of the $g=0$ and $g=\pm 1$ surfaces in the coordinate space defined by $(d_1,d_2,d_{\rm t})$ (pink surface: $g=0$; blue surface: $g=\pm 1$).
	}
	\label{fig:3d}
\end{figure}

Here, we plot the surfaces corresponding to $g=0$ and $g=\pm 1$ within the three-dimensional coordinate system defined by $(d_1,d_2,d_{\rm t})$, as shown in Fig.~\ref{fig:3d}. The pink surface represents the set of parameter combinations yielding $g=0$, while the blue surface delineates the boundaries of the stable region where $g=\pm 1$. It can be clearly observed that the stable region boundary ($g=\pm 1$) encloses the $g=0$ surface. This visualization helps to understand that if manufacturing or assembly tolerances lead to deviations in $(d_1,d_2,d_{\rm t})$, the actual operating point of the system might fall within the region bounded by $g=\pm 1$ (i.e., within the stable zone), or it might fall outside, leading to an unstable resonator.

\begin{figure}[t]
	\centering
	\includegraphics[width=3.2in]{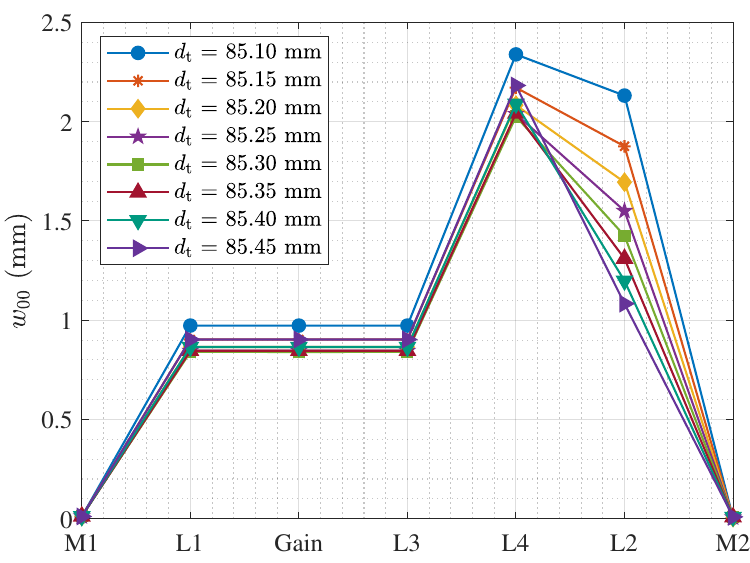}
	\caption{TEM$_{00}$ mode radius at different optical devices changes with the adjustment of $d_{\rm t}$ ($d_{\rm w}=6$~m).}
	\label{fig:W00dt}
\end{figure}

Figure~\ref{fig:W00dt} provides an intuitive visualization of the mode radius distribution at key optical components as $d_{\rm t}$ is adjusted. We see that the beam spot size is very small at mirrors M1 and M2. The beam is collimated between lenses L1 and L3 and is magnified by the telescope (formed by L3 and L4) as it propagates towards L2. Conversely, considering the reverse direction, the beam coming from free space is compressed upon passing through the telescope. Furthermore, as $d_{\rm t}$ is increased, the mode radius changes according to the behavior described in relation to Fig.~\ref{fig:W00g}. When $d_{\rm t}=85.05$~mm, the stability parameter $g$ approaches zero, resulting in the smallest mode radius at the gain medium. In practice, one might choose to set $d_{\rm t}$ slightly off the exact $g=0$ point, for example, at $d_{\rm t}=85.4$~mm, to achieve a small $w_{00}$ at L2 while maintaining the $w_{00}$ at the gain medium almost unchanged.

\subsection{The Maximum Acceptable Tolerance}
\label{sec:max_tolerance}

\begin{figure}[t]
	\centering
	\includegraphics[width=3.2in]{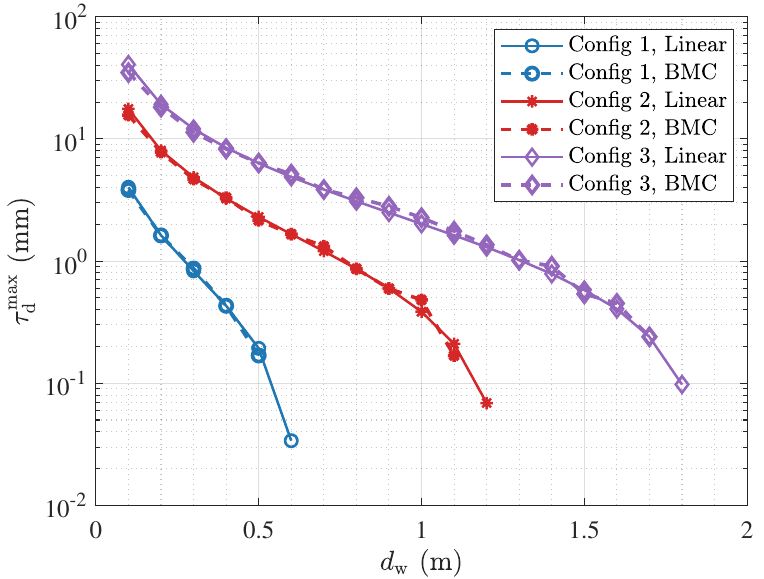}
	\caption{The maximum acceptable tolerance $\tau_{\rm d}^{\rm max}$ varies with working distance $d_{\rm w}$ obtained through linear appropriation and binary-search Monte Carlo simulation (BMC) algorithm\{Config 1:$(f_1,f_2,f_3,f_4)=(30,30,25,60)$~mm); $d_1=f_1$, $d_2=f_2$; $d_{\rm g}=f_1+f_3$; $d_{\rm t}$ is set upon $g=0$; Config 2 and Config 3 are twice and three times of Config 1, respectively\}.}
	\label{fig:tolerance}
\end{figure}

Figure~\ref{fig:tolerance} presents the maximum acceptable distance tolerance $\tau_{\rm d}^{\rm max}$, calculated using both the linear approximation method and the BMC algorithm, as a function of the working distance $d_{\rm w}$. Here the fixed tolerance $\tau_{\rm f}^*$ is set to be $1\%$ of the maximum focal length in each configuration. It is evident that  $\tau_{\rm d}^{\rm max}$ deteriorates rapidly as $d_{\rm w}$ increases. \mll Even when the overall focal lengths are scaled up by a factor of 3 compared to the basic configuration, the working distance limit where $\tau_{\rm d}^{\rm max}$ drops below $0.01$~mm improves by only approximately $1$~m,  remaining below the 2-m mark. \mrr This result provides a crucial guideline: \emph{it is not advisable} to rely solely on non-adjustable distance parameters set to their predetermined operating points for long $d_{\rm w}$, as typical manufacturing and assembly tolerances do not support such a design approach. Fortunately, Fig.~\ref{fig:Ddw} suggests a feasible strategy for achieving long $d_{\rm w}$. Specifically, for a kilometer-level SDC system, one should incorporate adjustability into components like the retroreflector or telescope, ensuring that the adjustment precision is within the acceptable range indicated by Fig.~\ref{fig:Ddw}. For instance, for a 10-m SDC, the required adjustment precision is on the order of $0.5$~mm, which is a feasible value. Even extending the working distance to 100 m remains achievable with larger focal length parameters, although the precision in need is very strict.

\subsection{Output Power and Working Distance Boundary}
\label{sec:output_distance_boundary}

\begin{figure}[t]
	\centering
	\includegraphics[width=3.2in]{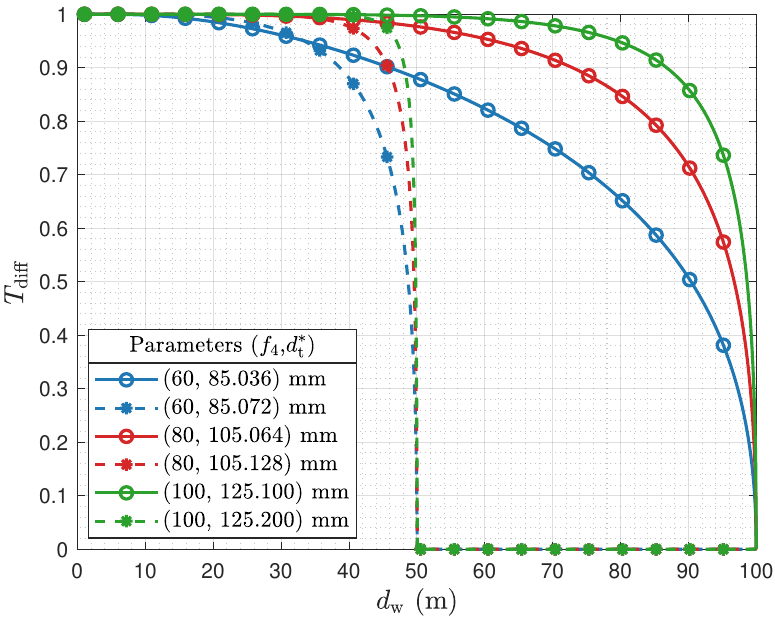}
	\caption{\mll Diffraction loss factor $T_{\rm diff}$ varies with working distance $d_{\rm w}$ ($g=0$).\mrr}
	\label{fig:Tdiffdw}
\end{figure}

The finite radius of the active gain medium is also a critical factor limiting the working distance $d_{\rm w}$. Since the aperture of the gain medium is typically smaller than other optical components in the cavity, it often contributes the dominant source of diffraction loss. As $d_{\rm w}$ increases, the $w_{00}$ within the gain medium also increases and eventually approaches the physical edge of the gain medium aperture. Because the resonator is highly sensitive to intracavity losses, this increasing diffraction loss causes the output power to deteriorate rapidly.

\begin{figure}[t]
	\centering
	\includegraphics[width=3.2in]{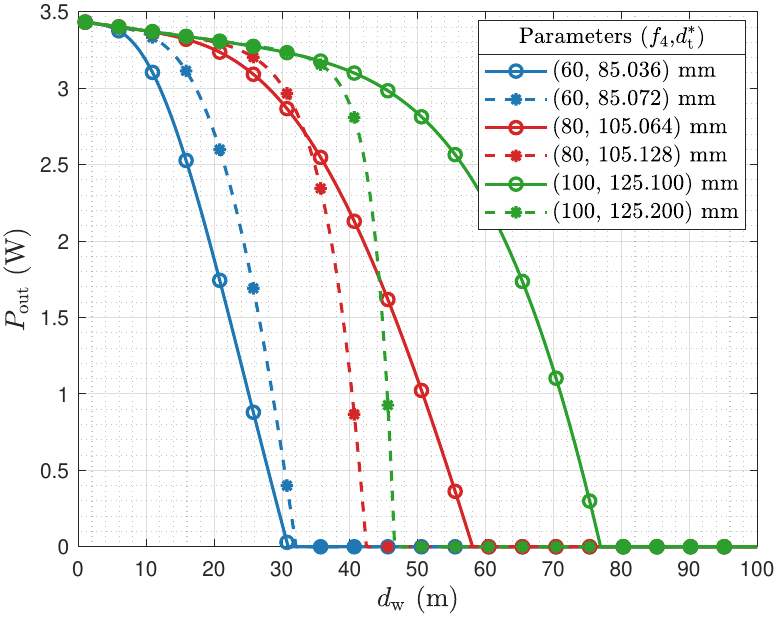}
	\caption{\mll Output power $P_{\rm out}$ varies with working distance $d_{\rm w}$ ($f_3=25$~mm, $g=0$).\mrr}
	\label{fig:Poutdw}
    \vspace{-0.3cm}
\end{figure}

As shown in Fig.~\ref{fig:Tdiffdw}, the diffraction loss factor $T_{\rm diff}$ remains close to 1 (implying minimal diffraction loss) over a significant range of working distances. However, it decreases sharply towards zero as $d_{\rm w}$ approaches a certain threshold. This threshold often corresponds to the stability parameter approaching the boundary $g=-1$, as can be inferred from Fig.~\ref{fig:G-D1DT}. In Fig.~\ref{fig:Tdiffdw}, $f_4$ is varied between $60$~mm, $80$~mm, and $100$~mm, while $f_3$ is fixed at $25$~mm. This configuration, where the telescope compresses the beam, provides efficient beam spot size reduction at the gain medium location. The parameter $d_{\rm t}^*$ here refers to the value of $d_{\rm t}$ optimized to achieve $g=0$ at a specific working distance $d_{\rm w}^*$. Two groups of configurations are compared: the dashed lines represent results where $d_{\rm t}^*$ is obtained for a preset working distance $d_{\rm w}^*=25$~m, while the solid lines correspond to $d_{\rm t}^*$ obtained for $d_{\rm w}^*=50$~mm. We can infer that if $d_{\rm t}^*$ is designed for a given $d_{\rm w}^*$, the diffraction loss remains low until $d_{\rm w}$ approximately reaches $2d_{\rm w}^*$, \mll the transmission distance can reach 100 m\mrr. From this figure, we also learn that incorporating a telescope generally improves the diffraction loss performance. Furthermore, using a telescope with larger magnification ($M_{\rm tel}=f_4/f_3$) allows the system to perform well over a longer range of working distances before diffraction loss becomes significant.

\begin{figure}[t]
	\centering
	\includegraphics[width=3.2in]{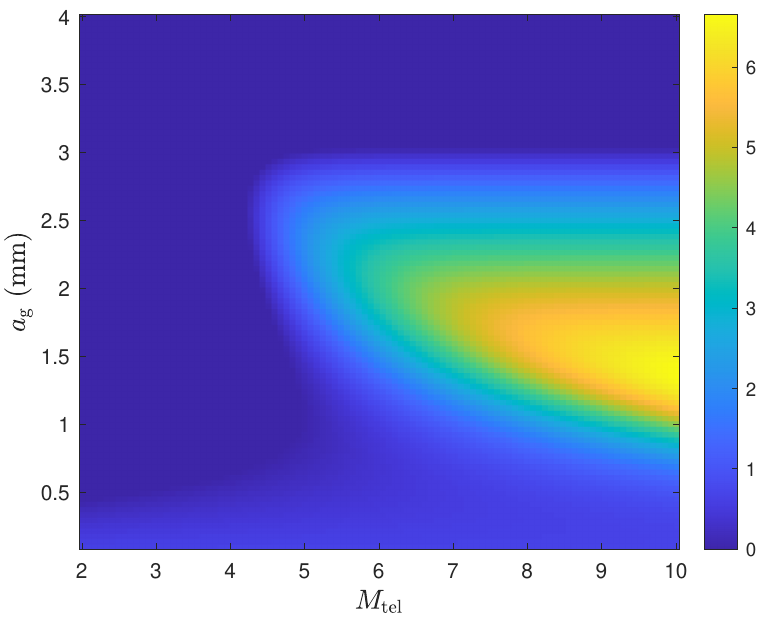}
	\caption{Output power $P_{\rm out}$ varies with gain medium radius $a_{\rm g}$ and telescope magnification $M_{\rm tel}$ ($d_{\rm w}=100$~m, $g=0$).}
	\label{fig:PouAgMtel}
\end{figure}

Figure~\ref{fig:Poutdw} supports the observations regarding diffraction loss and its impact on output power. \emph{Two main conclusions can be drawn}: Firstly, if a desired working distance $d_{\rm w}$ is targeted, $d_{\rm t}$ is recommenced to be adjusted to satisfy the $g=0$ condition at that distance; Secondly, to mitigate diffraction loss and improve performance at longer working distances, the magnification of the telescope should be increased. Simultaneously, an appropriate radius for the gain medium, $a_{\rm g}$, must be selected. \mll When $f_4 $ is 100 mm and $d_{\rm t}^*$ is 125.2 mm, the SDC system achieves a maximum transmission distance of 76.8 m. \mrr Figure~\ref{fig:PouAgMtel} intuitively demonstrates the combined impact of $a_{\rm g}$ and $M_{\rm tel}$ on the output power. For a designed working distance $d_{\rm w}$, a larger $M_{\rm tel}$ is generally preferred to maximize output power. The gain medium radius $a_{\rm g}$ requires careful selection, as both very large and very small values can lead to reduced output power.

\subsection{Experimental Setup and Results}

\begin{figure}[t]
	\centering
	\includegraphics[width=3.2in]{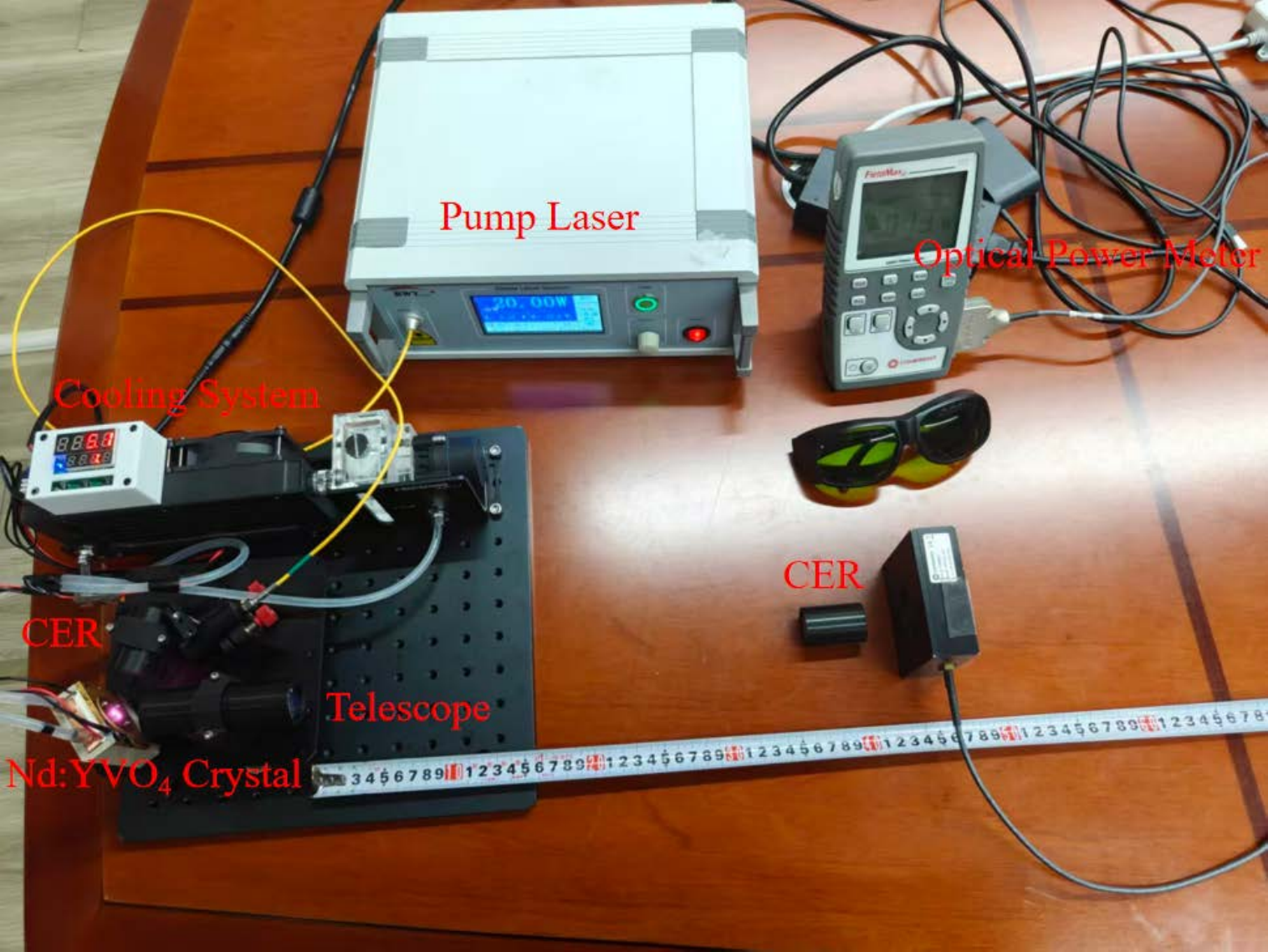}
	\caption{Schematic Diagram of the RBS Experiment}
    \label{fig:Experiment}
\end{figure}

\begin{figure}[t]
	\centering
	\includegraphics[width=3.2in]{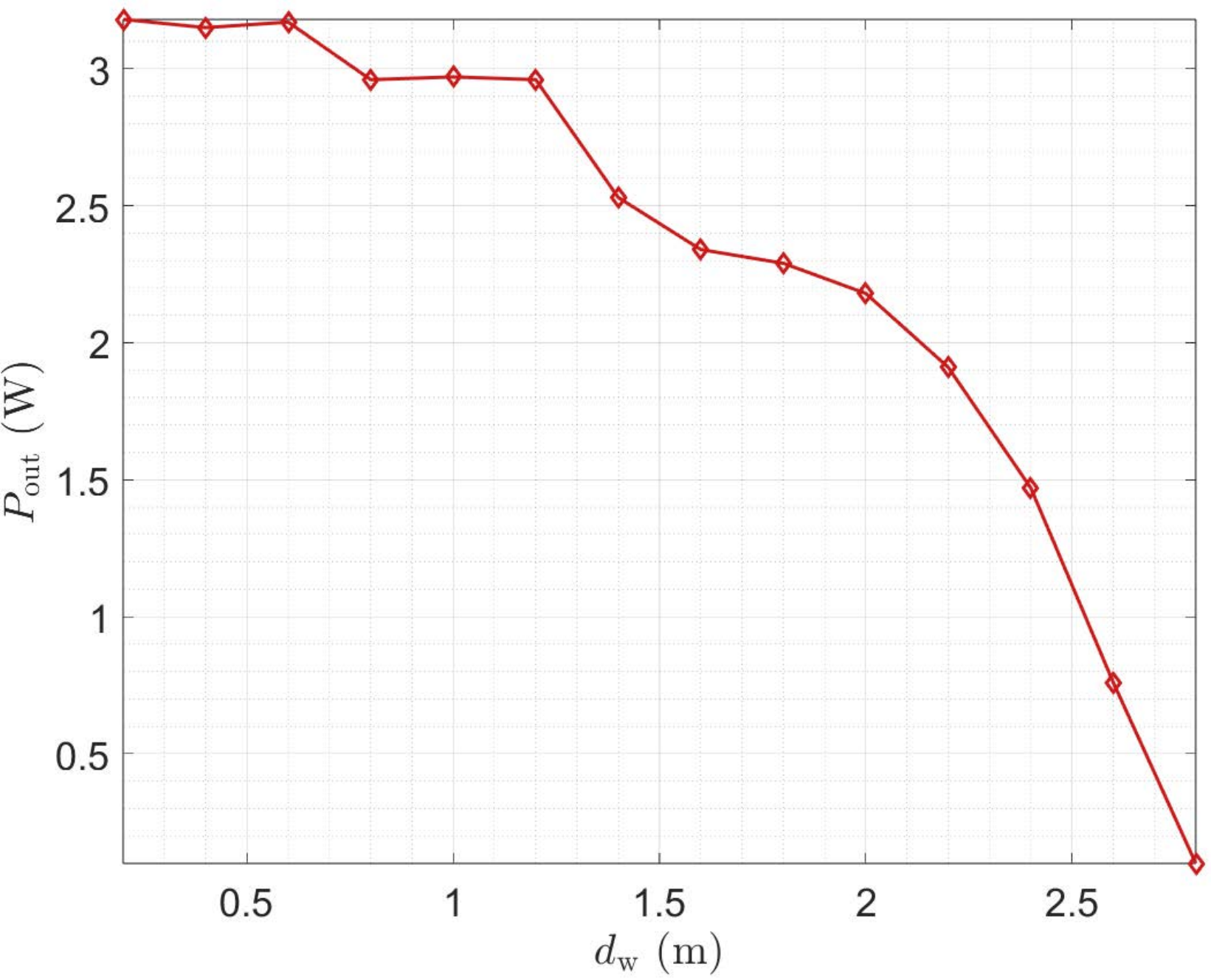}
	\caption{Output power $P_{\rm out}$ varies with transmission distance $d_{\rm w}$ in experiment.}
    \label{fig:Experimental results}
\end{figure}

\mll To verify the feasibility of adjusting the telescope-enhanced SDC system for greater working distance, we established the experimental platform as illustrated in Fig.~\ref{fig:Experiment}. Both the transmitter and receiver adopted a CER structure with a focal length of $f=30\ \text{mm}$ (see Config 1 in Fig.~\ref{fig:tolerance}). An Nd:YVO$_4$ crystal was employed as the gain medium, which absorbed a $20\ \text{W}$ pump laser at $880\ \text{nm}$ for optical amplification. Based on the operating mechanism of the laser resonator, a resonant beam at $1064\ \text{nm}$ was generated, with a telescope structure embedded in the resonator cavity. To prevent performance instability of the crystal caused by excessive temperature, a cooling system was attached to the crystal to maintain it at a relatively low temperature. The optical power received was detected by an optical power meter at the receiver end. To ensure experimental safety, the researchers wore OD4-level protective goggles throughout the experiment.\mrr

\mll The initial configuration restricted the beam initialization to within 0.5 m; however, following the adjustment of the stable region, the system demonstrated a capability of operating at distances greater than 2 m. Figure~\ref{fig:Experimental results} illustrates the variation characteristic of the output optical power \( P_{\text{out}} \) of the SDC system with the transmission distance \( d_{\text{w}} \). It can be observed that when  \( d_{\text{w}} \) is in the range of 0 -- 1.2 m, the output optical power \( P_{\text{out}} \) remains around 3 W, showing good stability; as \( d_{\text{w}} \) continues to increase, \( P_{\text{out}} \) exhibits a significant downward trend. When \( d_{\text{w}} = 2 \, \text{m} \), the output optical power drops to about 2.18 W, it can still achieve a transmission efficiency of 10.9\%. When \( d_{\text{w}} = 2.4 \, \text{m} \), the output optical power drops to about 1.47 W, at this point, the power has dropped to half its maximum value. Achieving resonance becomes difficult beyond 2.8 m, primarily because the required tuning precision exceeds the limits, thereby preventing the system from reaching a stable resonant state.\mrr

\subsection{Discussion}
\mll From the analysis of Fig. 9, it is evident that the maximum acceptable tolerance $\tau_{\rm d}^{\rm max}$ can only ensure the feasibility of the system within a $0.5$-m transmission range, which is insufficient to support long-distance SLIPT applications, but the stable region of the proposed SDC scheme exhibits superior adaptability, enabling the system to reliably exceed existing distance limitations. As further illustrated in Table~\ref{tab:tradeoffs}, the increase in transmission distance not only affects the output power of the laser but also leads to a notable variation in diffraction loss factor $T_{\rm diff}$ , these two parameters determine the overall efficiency and stability of the SLIPT system.\mrr

\begin{table}[htbp]
	\centering
	\caption{Parameter Trade-offs of Laser Cavity at Different Working Distances}
	\label{tab:tradeoffs}
	\begin{tabular}{c c c c}
		\hline
		\textbf{distance} & \textbf{$d_{\rm t}$ Stable Region Width} & \textbf{ $T_{\rm diff}$} & \textbf{Output Power} \\
		\textbf{(m)} & \textbf{(mm)} & \textbf{(\%)} & \textbf{(W)} \\
		\hline
		20 & 4.147 & 1 & 3.314 \\

		30 & 2.764 & 1 & 3.246 \\

		40 & 2.074 & 0.9991 & 3.112 \\
	
		50 & 1.659 & 0.9973 & 2.833 \\
	
        60 & 1.382 & 0.9919 & 2.264 \\
	
        70 & 1.185 & 0.9793 & 1.173 \\
	
        ... & ... & ... & ... \\
		\hline
	\end{tabular}
\end{table}

The insights derived from this analysis inform a design framework aimed at optimizing SDC system performance and extending its effective working distance. It is clear that the maximum acceptable tolerance $\tau_{\rm d}^{\rm max}$ in component manufacturing and assembly is a critical limiting factor for achieving long-distance resonant beam transmission. Due to inherent limitations in manufacturing processes, including the fabrication of optical tubes and lenses, it is challenging to produce fixed components with sufficient accuracy to maintain stability for working distances exceeding approximately 2 meters using fixed component intervals. 

Fortunately, our analysis reveals that the stable region width in terms of adjustable parameters like $d_{\rm t}$ is often significantly wider than the overall required tolerance for fixed systems, as shown in Fig.~\ref{fig:Ddw}. This implies that incorporating adjustability into either the $d_1$ or $d_{\rm t}$ parameter is a viable approach. In this scenario, the primary limitation shifts from manufacturing tolerance to the precision of the adjustment mechanism. For instance, as illustrated in Fig.~\ref{fig:Ddw}, designing a 10-meter SDC system would require an adjustment precision on the order of $0.5$~mm, which is technically feasible. Even extending the working distance to 100 meters remains achievable by utilizing larger parameter configurations and  ensuring higher adjustment precision. Finally, it is crucial to remember the importance of selecting an adequate combination of gain medium radius, $a_{\rm g}$, and telescope magnification, $M_{\rm tel}$, as this significantly impacts the efficiency and feasibility of long-range resonant beam formation.

\section{Conclusions}
\label{sec:con}

\mll This study presents a theoretical investigation and experimental verification into the properties and limitations of SDC system.\mrr We analyzed cavity stability, beam parameters, and output power, with particular emphasis on the influence of manufacturing and assembly tolerances on system stability. We developed and applied two complementary methods: a binary-search-based Monte Carlo simulation and a linear approximation method to determine the maximum acceptable distance tolerance. Numerical results demonstrate that achieving stable operation over long working distances is severely constrained by typical tolerances, as the stable region width decreases with the working distances. Consequently, incorporating adjustability, particularly for parameters like $d_{\rm t}$ (lens-to-lens interval of the telescope), is crucial to ensure long-distance SDC operation. Furthermore, diffraction loss at the gain medium aperture emerges as a primary factor limiting output power at extended distances, emphasizing the need to jointly optimize the active gain medium radius and the magnification of the telescope. \mll Experimental system verified the limitation of fabrication tolerance can only ensure working distance less than $0.5$~m, and by adjusting the stable region in assembly the distance can reach $2.8$~m. This work offers practical guidelines for the design and implementation of stable SDC systems, enabling long-range resonant beam formation to address the demands of remote sensor networks, dynamically charged/communicated mobile IoT devices, and EMI-sensitive IoT deployments.\mrr





\ifCLASSOPTIONcaptionsoff
  \newpage
\fi

\bibliographystyle{IEEETran}
\small

\bibliography{mybib}

\end{document}